\documentclass[fleqn,usenatbib]{mnras}

\usepackage{mathptmx}
\usepackage{txfonts}

\usepackage[T1]{fontenc}

\DeclareRobustCommand{\VAN}[3]{#2}
\let\VANthebibliography\thebibliography
\def\thebibliography{\DeclareRobustCommand{\VAN}[3]{##3}\VANthebibliography}

\usepackage{graphicx}	
\usepackage{amssymb}	
\usepackage{epsfig}

\newcommand{\msun}{$\mathrm{M_{\sun}}$}
\newcommand{\rsun}{$\mathrm{R_{\sun}}$}
\newcommand{\teff}{$T_{\rm eff}$}

\title[Two CoRoT solar analogues]{Modeling of two CoRoT solar analogues constrained by seismic and spectroscopic analysis}

\author[M. Castro et al.]{
M. Castro$^{1}$\thanks{email: mcastro@fisica.ufrn.br},
F. Baudin$^{2}$,
O. Benomar$^{3,4}$,
R. Samadi$^{5}$,
T. Morel$^{6}$,
\newauthor{
C. Barban$^{5}$,
J. D. do Nascimento Jr.$^{1,2,7}$,
Y. Lebreton$^{5,8}$,
}
\newauthor{
P. Boumier$^{2}$,
J. P. Marques$^{2}$,
and J. S. da Costa$^{9}$
}
\\
$^{1}$Departamento de F\'isica, Universidade Federal do Rio Grande do Norte, CEP: 59072-970 Natal, RN, Brazil\\
$^{2}$Universit\'e Paris-Sud, CNRS, Institut d'Astrophysique Spatiale, UMR 8617, 91405, Orsay Cedex, France\\
$^{3}$Center for Space Science, NYUAD Institute, New York University Abu Dhabi, PO Box 129188, Abu Dhabi, UAE\\
$^{4}$Division of Solar and Plasma Astrophysics, NAOJ, Mitaka, Tokyo, Japan\\
$^{5}$LESIA, Observatoire de Paris, Universit\'e PSL, CNRS, Sorbonne Universit\'e, Universit\'e de Paris, 5 place Jules Janssen, 92195 Meudon, France\\
$^{6}$Space sciences, Technologies and Astrophysics Research (STAR) Institute, Universit\'e de Li\`ege, Quartier Agora, All\'ee du 6 Ao\^ut 19c, B\^at. B5C, B4000-Li\`ege, Belgium\\
$^{7}$Harvard-Smithsonian Center for Astrophysics, Cambridge, MA 02138, USA\\
$^{8}$Universit\'e Rennes, CNRS, IPR (Institut de Physique de Rennes) - UMR 6251, F-35000 Rennes, France\\
$^{9}$Escola de Ci{\^e}ncia e Tecnologia, Universidade Federal do Rio Grande do Norte, CEP: 59072-970 Natal, RN, Brazil
}

\date{Accepted XXX. Received YYY; in original form ZZZ}

\pubyear{2021}
\hypersetup{final}
\begin{document} 
\label{firstpage}
\pagerange{\pageref{firstpage}--\pageref{lastpage}}
\maketitle

\begin{abstract}
Solar analogues are important stars to study for understanding the properties of the Sun. Evolutionary modeling, combined with seismic and spectroscopic analysis, becomes a powerful method to characterize stellar intrinsic parameters, such as mass, radius, metallicity and age. However, these characteristics, relevant for other aspects of astrophysics or exoplanetary system physics for example, are difficult to obtain with a high precision and/or accuracy. The goal of this study is to characterize the two solar analogues HD42618 and HD43587, observed by CoRoT. In particular, we aim to infer precise mass, radius, and age, using evolutionary modeling constrained by spectroscopic, photometric, and seismic analysis. These stars show evidences of being older than the Sun but with a relatively large lithium abundance. We present the seismic analysis of HD42618, and the modeling of the two solar analogs HD42618 and HD43587 using the CESTAM stellar evolution code. Models were computed to reproduce the spectroscopic (effective temperature and metallicity) and seismic (mode frequencies) data, and the luminosity of the stars, based on Gaia parallaxes. We infer very similar values of mass and radius for both stars compared to the literature, within the uncertainties, and reproduce correctly the seismic constraints.  For HD42618, the modeling shows it is slightly less massive and older than the Sun. For HD43587, it confirms it is more massive and older than the Sun, in agreement with previous results. The use of chemical clocks improves the reliability of our age estimates.
\end{abstract}
   



\section{Introduction}
\label{sec:intro}

The characterization of solar analogues and solar twins is a powerful and promising approach to better understand stellar evolution, and more specifically the evolution of the Sun itself and the influence of leading parameters such as stellar mass and metallicity. The canonical differentiation of solar analogues with respect to solar twins comes from \citet{cayrel81} and \citet{cayrel96}. These authors described a solar twin as a star spectroscopically and photometrically indistinguishable from the Sun, within observational uncertainties, while solar analogues present up to a 10\% difference in their radius and mass and a difference of less than $\pm 0.2$ dex in metallicity when directly compared to the Sun (\citealt{melendez10}; \citealt{beck17}). Among stellar properties, age is not yet taken into account in these definitions due to intrinsic difficulties to estimate it.

The  number of solar analogue stars has increased along the last decade, showing slightly different properties such as rotation period, age or magnetic activity \citep{garcia14,baumann10,schrijver&zwaan08} particularly in the last years thanks to the remarkable quality of the continuous photometric observations obtained by the CoRoT \citep{baglin06}, Kepler \citep{borucki10}, and on-going TESS \citep{ricker15} space missions, as well as Gaia satellite measurements \citep{gaia16,gaia18,gaia20}. These observatories provide time series of the stellar brightness, as light curves, for tens of thousands of stars. These sets of data, and the associated signal processing techniques, allowed the measurement of fundamental parameters and acoustic oscillations for hundreds of solar-like oscillating stars \citep[e.g.][]{chaplin14}.

In addition to the spectroscopic and photometric standard analyses, asteroseismology is a major tool to better define and study solar analogues and twins \citep{bazot12}. \citet{appourchaux08} and \citet{benomar09b} present one of the first asteroseismic analysis on a F5V CoRoT star showing Sun-like oscillations, HD49933, and extract several \textit{p}-mode frequencies, the large frequency spacing, the frequency of maximum amplitude of the modes, and the mean rotational frequency splitting. \citet{piau09} compared these results to stellar models to estimate the impact of input physics on classical and seismic parameters. They pointed out that diffusion and rotation-induced mixing have to be included in the models to achieve reliable mass and age estimates. However, they did not aim to find the best model that fit the observational constraints to estimate mass and age. \citet{lebreton&goupil14} performed a very detailed modeling of another CoRoT star, HD\,52265, a metal-rich G0V star, more massive than the Sun. They explored many of the parameters and approaches that can influence the results of modeling. Another example of astronomical analysis of a solar analogue is the characterization of 16 Cyg A \& B, based on \textit{Kepler} observations \citep[e.g.][]{metcalfe12,donascimento14,davies15,bazot19}. \citet{donascimento14} complemented the light curve analysis by comparing with stellar evolution models. The brightest solar twin, 18 Sco, has been studied by \citet{bazot18}, which used spectrophotometric, interferometric and asteroseismic data to constrain stellar evolution models and estimate physical characteristics. They reached a precision of 6\% on the mass and $X_{0}$, 9\% on $Y_{0}$, and 35\% on the mixing-length parameter. Recently, \citet{nsamba21}, using asteroseismic inferences, quantified the effect of the treatment of the initial helium abundance on the systematic uncertainties on the inferred stellar parameters, such as radius, mass, and age, in stellar model grids.\

\citet{morel13} obtained high-resolution spectroscopy of two bright solar analogues CoRoT targets, HD42618 and HD43587, with the HARPS spectrograph. Thanks to the relatively high brightness of these stars and a S/N ratio of about 300, the exploitation of these observations is made easier and more robust. They presented atmospheric parameters and chemical composition of both stars, precisely determined using a fully differential analysis with respect to the Sun. Although both stars are confirmed to be solar analogues, they found differences in the surface abundance of lithium, which could be explained by different mixing efficiencies in their interiors. They pointed out that these results should set constraints on theoretical modeling of the internal structures and solar-like oscillations of these stars. \citet{boumier14} carried out a seismic analysis of HD43587. They extracted 26 \textit{p}-mode frequencies with radial degrees \textit{l} = 0, 1, and 2, and from modeling with the stellar evolution code Cesam2k \citep{morel&lebreton08} and the LOSC adiabatic pulsation code \citep{scuflaire08}, they determined that HD43587 seems to be slightly more massive and older than the Sun.
 
In this context, we propose to deepen the combined seismic and spectroscopic analysis of bright stars by studying the two CoRoT solar analogues, HD42618 and HD43587, using the stellar evolution code CESTAM \citep{marques13}. For each star, we find the best-fitted model that accounts for spectroscopic, photometric and asteroseismic observations, as explained in Sect. \ref{sec:Models}, to infer a mass, radius and age estimate.

Our paper is organized as follows: in Sect. \ref{sec:Obs} we present relevant observational informations about the two CoRoT targets HD43587 and HD42618. In Sect. \ref{sec:Models}, we present the stellar evolution code CESTAM, used to model both stars, as well as the calibration and optimization procedure. In Sect. \ref{sec:results} we present our modeling results. Finally, we give our conclusions in Sect. \ref{sec:conclusions}.

\section{Two CoRoT solar analogue stars}
\label{sec:Obs}

We studied here two targets of the CoRoT mission, HD42618 and HD43587, observed through the so-called \textit{seismic channel} aiming at bright stars \citep{ollivier16}, allowing precise spectroscopic observations and thus a combined seismic and spectroscopic analysis. These two targets are the closest to solar characteristics among the CoRoT sample of bright stars. As more spectroscopic data will become available, other stars such as Kepler (however fainter) or TESS targets (with generally shorter time series) could be included in future works.

\subsection{HD43587}

HD43587, a G0V star observed by CoRoT for 145 days, has been observed with the high resolution spectrograph HARPS at La Silla in December 2010-January 2011, to reach a S/N ratio higher than 300. The analysis of the spectroscopic data is presented in \citet{morel13}, from which we retain the following spectroscopic characteristics: effective temperature \teff = $5947 \pm 17$ K, and metallicity [Fe/H]$= -0.02 \pm 0.02$.
These small uncertainties are the result of a differential analysis of HD43587 with respect to the Sun, both stars having very similar parameters, which is expected to minimize systematic errors \citep[see][for more details]{morel13}. Differential analysis of solar analogues with respect to the Sun in the literature quote similar (actually even [much] smaller) uncertainties \citep[see, e.g., Table 2 of][]{spina18}.

A first analysis of the seismic data of HD43587 has been made by \citet{boumier14}. In the following, we used the oscillations frequencies they measured \citep[see Table 1 in][]{boumier14}. The authors derived from the seismic data a mass and a radius slightly larger than the solar values ($M = 1.04 \pm 0.01$ \msun, $R = 1.19$ \rsun) and an age larger than the solar one, $5.60 \pm 0.16$ Gyr, in apparent contradiction with its high lithium abundance \citep[$A(\mathrm{Li}) = 2.05 \pm 0.05$,][]{morel13}, which is an order of magnitude larger than solar abundance. Such enrichment is not expected for this type of star at that age \citep{melendez10}.

\subsection{HD42618}

Our second target is HD42618, another CoRoT target (a G4V star), observed twice for 79 and 94 days of duty cycle, and that has been spectroscopically characterized from several different observations. A preliminary seismic analysis was done by \citet{barban13}.

For the seismic analysis of HD42618, we used the time series provided by the CoRoT public archive\footnote{\url{http://idoc-corot.ias.u-psud.fr}}. It corresponds to the so-called N2 data that are corrected from various instrumental effects \citep{chaintreuil16,ollivier16}. The star was observed during CoRoT periods LRa04 (28 September 2010 - 16 December 2010) and LRa05 (17 December 2010 - 22 March 2011), corresponding to a total observation duration of 184 days. The duty cycle is of about $95\%$ so that gaps in the time series are expected to have marginal impact on the data analysis. The light curve is prepared using the same method as in \cite{appourchaux08} and is analyzed on the Fourier space after computing its power spectrum using the fast Fourier transform.

\begin{figure}
  \begin{center}
  	\epsfig{figure=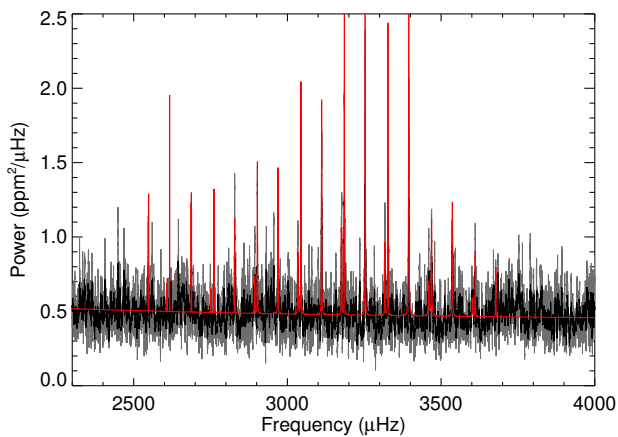, angle=0, width=8cm}
   \end{center}
\caption{Power spectrum of HD42618 smoothed using a box-car of width $1\mu$Hz (gray) and $3\mu$Hz (black).  Superimposed is shown the best fit (red).}\label{fig:psf}
\end{figure}

\begin{figure}
  \begin{center}
  	\epsfig{figure=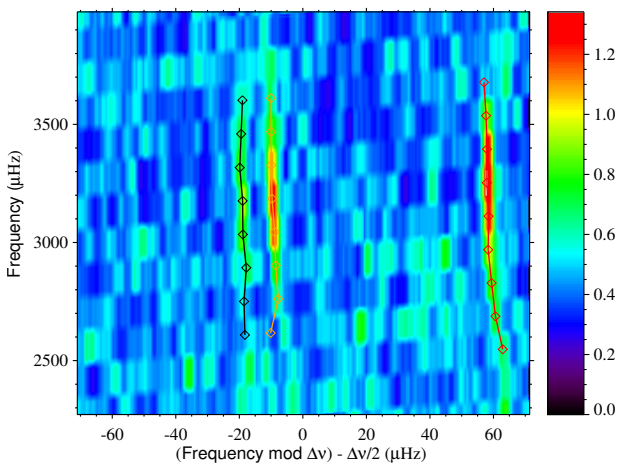, angle=0, width=8cm}
   \end{center}
\caption{Echelle diagram for HD42618. Frequencies for the best fit are shown in orange ($l=0$), red ($l=1$) and black ($l=2$).}\label{fig:ED}
\end{figure}

Although of weak amplitude, the \textit{p} modes of HD42618 are apparent on the power spectrum (Fig.\,\ref{fig:psf}). The mode identification is performed in the \'echelle diagram (Fig.\,\ref{fig:ED}), a concept introduced by \cite{grec83}. The \'echelle diagram shows two clear ridges associated to the $l=0$ and $l=1$ modes and a fainter one due to $l=2$ modes. Modes of degree greater than $l=2$ are not visible due to their low amplitudes.

\begin{table*}
\caption{Measured mode frequency $\nu$, height $H$, width $\Gamma$ for modes of degree $l=0,1,2$ for HD42618. Symmetric uncertainties $e_\nu$ are given for frequencies, contrary to uncertainties on the other parameters that follow the format $e^+_X$ and $e^-_X$, $X$ being the parameter.}
\centering
\begin{tabular}{c|c|c|c|c|c|c|c|c }
$l$  & $\nu$ ($\mu$Hz) & $e_\nu$ ($\mu$Hz) & $H$ (ppm$^2/\mu$Hz)  & $e^+_H$   &  $e^-_H$  & $\Gamma$ ($\mu$Hz) &  $e^+_\Gamma$  & $e^-_\Gamma$ \\ \hline \hline
         0 &  2616.85   &     4.15  &     1.49    &    1.72   &     0.84   &   0.06     &     0.16   &       0.05  \\  
	0 &   2761.25   &    1.10   &   0.83    &     0.53    &    0.31   &      0.26  &      0.35   &      0.16  \\ 
	0 &   2902.45   &   0.17    &   1.02      &   0.54    &    0.33   &     0.77   &     0.53    &     0.31 \\
	0 &  3044.13    &    0.16   &   1.56    &     0.52    &    0.39   &     0.92   &     0.29    &     0.24 \\
	0 &   3185.32   &     0.12  &   2.52    &     2.07    &    0.70   &     0.60   &     0.22    &     0.24 \\
	0 &  3327.06    &    0.24   &   1.97    &     0.75    &    0.61   &     0.79   &     0.42    &     0.21\\
	0  &  3469.00   &    0.37   &   0.65    &     0.26    &    0.18   &     1.75   &     0.57    &     0.59 \\
	0  &  3610.99   &    0.44   &   0.44    &     0.50    &    0.34   &     0.90   &     3.77    &     0.46 \\
	1 &   2547.92   &    0.63   &   2.60    &     3.22    &    1.95   &     0.14   &     0.52    &     0.10 \\
	1 &   2687.62   &    0.27   &   2.24     &    2.57    &    1.25   &     0.06   &     0.16    &     0.05  \\
	1  &  2828.41   &    0.23   &  1.24     &    0.80     &   0.46    &    0.26    &     0.35    &     0.16  \\
	1 &   2969.33   &    0.29   &  1.52     &    0.81     &   0.50    &    0.77    &     0.53    &     0.31  \\
	1 &   3111.44    &    0.17   &    2.33     &   0.78    &    0.58   &     0.92    &      0.29  &       0.24  \\
	1 &   3252.91   &     0.18  &   3.77    &     3.10    &    1.05    &     0.60  &       0.22  &      0.24  \\
	1 &   3394.96   &    0.14   &   2.95     &    1.12    &    0.92   &     0.79   &      0.42   &     0.21  \\
	1 &   3536.67   &    0.41   &   0.98     &    0.38    &    0.27   &     1.75   &      0.57   &     0.59  \\
	1 &   3678.0     &   1.50    &   0.66    &     0.74    &    0.51   &     0.90   &      3.77   &     0.46 \\
	2 &   2608.75   &   4.15    &  0.79      &   0.91     &   0.45    &   0.06     &    0.16     &   0.05 \\
	2 &   2750.48   &    1.69   &  0.44     &    0.29     &   0.17    &   0.26     &    0.35     &   0.16 \\
	2 &   2893.19   &    0.59   &  0.54     &    0.29     &   0.18    &  0.77      &   0.53      &  0.31  \\
	2 &   3034.12   &    0.41   &  0.82     &    0.28     &   0.21    &  0.92      &   0.29      &  0.24  \\
	2  &  3176.04   &    0.17   &  1.33       &   1.10    &    0.38   &    0.60    &    0.22     &   0.24  \\
	2 &   3317.13   &    0.41   &  1.04       &   0.40    &    0.33   &    0.79    &    0.42     &   0.21 \\
	2  &  3459.57   &    0.66   &  0.34       &   0.14    &    0.10   &    1.75    &    0.57     &   0.59 \\
	2  &  3601.98   &    1.45   &  0.23      &    0.27    &    0.18   &     0.90   &     3.77    &    0.46  \\
\hline
\end{tabular} 
\label{tab:1}
\end{table*}

In order to reliably extract pulsations characteristics, we perform a Bayesian analysis. First, we measure the global properties of the acoustic modes using the pipeline described in \cite{benomar12}. Mode amplitudes follow a bell-shaped function often modeled as a Gaussian, over the noise background. Here, we fit such a model, with the noise background being described by the sum of two power-laws \citep{harvey85} plus a white noise.  This allows us to measure the frequency at maximum amplitude $\nu_{max}$ which relates to the mass, radius and effective temperature of the star \cite[e.g.][]{huber11}. We found $\nu_{max} = 3157 \pm 46 \,\mu$Hz. This is strikingly similar to the solar value ($\nu_{max, \odot} = 3090 \pm 30 \,\mu$Hz), as per reported in the literature \citep{huber11}.

Acoustic frequencies of high order and low degree $(n \gg 1,\, l \sim 1)$ are nearly equally spaced and separated on average by a frequency spacing $\Delta\nu$. The spacing is related to the sound velocity inside the star by $\Delta\nu = (2 \int_0^R dr/c(r))^{-1}$, which is proportional to the mean stellar density $\bar{\rho}$. Because the solar density $\bar{\rho}_{\odot} = (1.4060 \pm 0.0005) \times 10^3$ kg.m$^{-3}$ and frequency spacing $\Delta\nu_{\odot} = 135.2 \pm 0.45 \,\mu$Hz \citep{huber11} are accurately known,  it is possible to  reliably estimate the mean density of any Sun-like star by the scaling relation, $\bar{\rho} = \bar{\rho}_{\odot}(\Delta\nu/\Delta\nu_\odot)^2$. For HD42618, using the EACF method \citep[Envelope Auto Correlation Function,][]{mosser09}, we found $\Delta\nu = 141.2 \pm 0.6\,\mu$Hz, which gives $\bar{\rho} = (1.554 \pm 0.025) \times 10^3$ kg.cm$^{-3}$, a density slightly higher than the Sun.

The precise determination of individual pulsation properties, and in particular the frequencies, is done in a similar fashion to, e.g., \citet{appourchaux08}, \citet{benomar09b}, \citet{handberg11}, \citet{ballot11}, and \citet{benomar14}. More specifically, we use the MCMC sampling algorithm from \cite{benomar09a}. The power spectrum is modeled as a sum of Lorentzian profiles, with frequency, height, width, rotational splitting and the stellar inclination angle as free parameters. The noise background function is again a sum of power laws. The Table\,\ref{tab:1} shows the frequencies, widths and heights of the modes for the best fit using the median as statistical indicator, along with the $1\sigma$ uncertainties. Due to the low spectral resolution $r=0.066\,\mu$Hz and to important correlations between the rotational splitting $\delta\nu$ and the stellar inclination $i$, it is difficult to measure individually these parameters for that star. However, the projected rotation $\delta\nu.\sin(i)=0.36 \pm 0.08 \,\mu$Hz is well constrained. The large separation derived from the frequency list, $\Delta\nu = 142.0 \pm 0.6\,\mu$Hz is consistent with the result from the EACF method.

\citet{morel13} performed a similar spectroscopic differential analysis as for HD43587 based on HARPS observations and derived an effective temperature $T_{\rm eff}=5765 \pm 17$\,K,  and  a metallicity [Fe/H]$= -0.10 \pm 0.02$. This star has also been observed by several other authors: \citet{fulton16} derived $T_{\rm eff} = 5747 \pm 44$\,K and [Fe/H]$= -0.11 \pm 0.03$ from HIRES observations at the Keck telescope. These authors also claim the presence of a neptunian planet around HD42618. \citet{mahdi16} used ELODIE measurements and a differential analysis to derive $T_{\rm eff}= 5766 \pm 13$\,K and [Fe/H]$= -0.09 \pm 0.01$. HD42618 was also analyzed by \citet{ramirez14} who found very similar results ($T_{\rm eff} = 5758 \pm 5$\,K and [Fe/H] $= -0.096 \pm 0.005$) , and more recently by \citet{spina18} ($T_{\rm eff} = 5762 \pm 3$\,K and [Fe/H] $= -0.112 \pm 0.003$) . These independent results show  an excellent  agreement, in particular those based on a differential analysis giving confidence about effective temperature and metallicity. In order to have spectroscopic data homogeneous with those of HD43587,  we retain the values derived by \citet{morel13}.

\subsection{Luminosity estimate of both stars}

To compare the observational data to the models, we need to estimate the luminosity of these two stars. For both of them, extinction was neglected due to their small distance (lower than 25 pc). For the luminosity calculation, we made use of the Gaia DR2 parallaxes, that are available for both stars \citep{gaia18,luri18}. For HD43587, we used the V-magnitude given in the SIMBAD database $V = 5.700 \pm 0.009$ \citep{oja91}, the Gaia DR2 parallax $\pi = 51.803 \pm 0.111\,\mathrm{mas}$ \citep{gaia18}, and the bolometric correction computed according to \citet{vandenberg&clem03} $BC = -0.048 \pm 0.006$. We used the prescription of \citet{zinn19} to calculate the magnitude-dependent zero-point offset of Gaia DR2 parallaxes due to instrumental effects, in particular basic-angle variations \citep{lindegren18}. We found an offset of $0.063 \pm 0.014$ mas, to be added to the original Gaia DR2 parallax. We obtained for the luminosity $L/L_{\odot} = 1.605 \pm 0.037$.

We used the same method for HD42618, using the V-magnitude in the SIMBAD database $V = 6.839 \pm 0.012$ \citep{koen10}. Using the Gaia DR2 parallax $\pi = 41.063 \pm 0.042\,\mathrm{mas}$ \citep{gaia18}, with a zero-point offset of $0.059 \pm 0.013$ mas, and the bolometric correction $BC = 0.077 \pm 0.008$, we found $L/L_{\odot} = 0.918 \pm 0.012$.

For HD42618, a significant discrepancy appears between the Hipparcos parallax \citep[\(\pi = 42.55 \pm 0.55\,\mathrm{mas}\),][]{vanleeuwen07} and the Gaia DR2 result. Differences between parallaxes from Hipparcos and Gaia are expected and can be positive or negative (based on few examples drawn from CoRoT targets) and Gaia error bars are always smaller by at least a factor of 2. Both parallaxes are generally consistent due to the larger Hipparcos error bars. However, in the case of HD42618, parallaxes, and thus derived luminosities, which is $L/L_{\odot} = 0.858 \pm 0.025$ using Hipparcos parallax, are not consistent within 1$\sigma$ error bars.

\section{Stellar evolutionary models}
\label{sec:Models}

We present in this section the stellar evolution code CESTAM \citep{marques13, morel&lebreton08, morel97} used to model the stars considered. 

The input physics used in the models are the OPAL05 equation of state \citep{rogers&nayfonov02} and the OPAL opacities \citep{iglesias&rogers96}, complemented,  at temperatures lower than $10^4 {\rm K}$, by the WICHITA opacities \citep{ferguson05}. Nuclear reaction rates were obtained using the NACRE compilation \citep{angulo99}, except for the \(^{14}\mathrm{N}(p,\gamma)^{15}\mathrm{O}\) for which we used the rates derived by \citet{formicola04}.

Convective instability was determined according to the Schwarzschild criterion. In convective zones, the temperature gradient was computed using the so-called CGM description following \citet{canuto96}. We adopted the solar mixture of \citet{grevesse&noels93}. We also computed models based on the mixing-length convection treatment of \citet{bohmvitense58} and \citet{asplund09} solar mixture, but they did not fit observations satisfactorily, and the adopted convection treatment and initial chemical mixture allowed a better agreement of the computed oscillation frequencies with the observed ones, from the optimization described hereafter.

Following the formalism of \citet{michaud&profitt93}, models were computed including microscopic diffusion of helium and heavy elements by gravitational settling, thermal and concentration diffusion but no radiative levitation. CESTAM includes transport of angular momentum by meridional currents and shear turbulence according to \citet{zahn92}. However, it is well known that this prescription does not reproduce the observed rotation profile of the Sun and red giants. Moreover, it also fails to reproduce the observed lithium abundance of the Sun. For this reason, we did not follow the lithium abundance evolution in CESTAM models. 

Oscillation frequencies were computed using the ADIPLS adiabatic oscillation code \citep{jcd08}.

A minimization algorithm, called OSM\footnote{Optimal Stellar Models, see https://pypi.org/project/osm/}, based on the Levenberg-Marquardt method, was used in order to determine the optimum CESTAM model matching the observational constraints. In this algorithm, some model parameters are allowed to vary. In the present work, the model parameters adjusted in order to fit observational constraints were:\\
- $M$, the stellar mass\\
- $A$, the age;\\
- $\alpha_{\rm CGM}$, the constant used in the CGM description of the convection;\\
- $Y_0$, the initial helium abundance;\\
- $Z_0$, the initial metallicity.\\
In addition, the surface effects affecting the mode frequencies (and the frequency separations listed below) are taken into account following the prescription proposed by \citet{kjeldsen08}, which two parameters, $a$ and $b$, were fitted following:
\begin{equation}
    \nu_{\mathrm{obs}}-\nu_{\mathrm{mod}}=a\left(\frac{\nu_{\mathrm{obs}}}{\nu_0}\right)^b
\end{equation}
here $\nu_{\mathrm{obs}}$ and $\nu_{\mathrm{mod}}$ are the observed and modeled frequencies and $\nu_0$ a reference frequency.
The observational constraints included global characteristics of the star plus seismic constraints:\\
- $T_{\rm eff}$, the effective temperature;\\
- [Fe/H], the observable that is a proxy of the surface metallicity\\
- $L$, the luminosity;\\
- $\nu_{n,\ell}$, the individual frequencies of all the observed modes;\\
- $\Delta\nu_0$, the individual seismic large separations for $\ell=0$: $\Delta\nu=\nu_{n,0}-\nu_{n-1,0}$;\\
- $r_{01} = \delta\nu_{01}/\Delta\nu_1$, the ratio of the second individual differences between $\ell=0$ and $\ell=1$ modes \citep[see][]{roxburgh&vorontsov03} normalized by the large separation of $\ell=1$ modes $\Delta\nu_1$, with: $\delta\nu_{01}=(\nu_{n-1,0}-4\nu_{n-1,1}+6\nu_{n,0}-4\nu_{n,1}+\nu_{n+1,0})/8$.\\
- $r_{02} = \delta\nu_{02}/\Delta\nu$, the individual seismic small separations: $\delta\nu_{02}=\nu_{n,l=0}-\nu_{n-1,l=2}$ normalised by the mean large separation of $\ell$ = 0, 1 and 2 modes.\\

The free model parameters listed above are adjusted in order to minimize the differences between computed and observed constraints (also listed above) by finding the lowest value of the $\chi^2$ between them. Using this approach, uncertainties on parameters are computed for fitted parameters using the Hessian matrix. The correlation between the fitted parameters is taken into account through the covariance matrix \citep[following][]{miglio05}. However, some characteristics of the star, such as the radius or the effective temperature for example, are output of the optimum model. They cannot be associated to an uncertainty since they are not adjusted during the minimizing process.\\

\section{Results}
\label{sec:results}

In this section, we present the results of our calculations for both stars. We then compare to the spectroscopic and seismic inferences from the literature.\\

\subsection{HD43587}

\begin{figure}
 \centering
 \includegraphics[width=9cm,height=9cm]{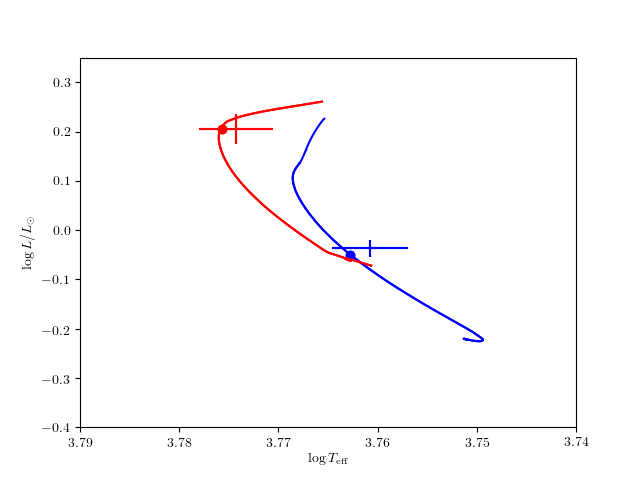} 
 \caption{HR diagram for the stars HD43587 (red) and HD42618 (blue). Continuous lines are the evolution tracks of the best-fitted CESTAM models reproducing spectroscopic, photometric and seismic observations, represented by the filled circles (see Sect. \ref{sec:results}). Crosses represent the observed values and their associated 3$\sigma$ error bars as described in Sect.\ref{sec:Obs}.}
 \label{fig:HRdiag}
\end{figure}

\begin{figure*}
 \centering
 \includegraphics[width=14cm,height=6cm,angle=0]{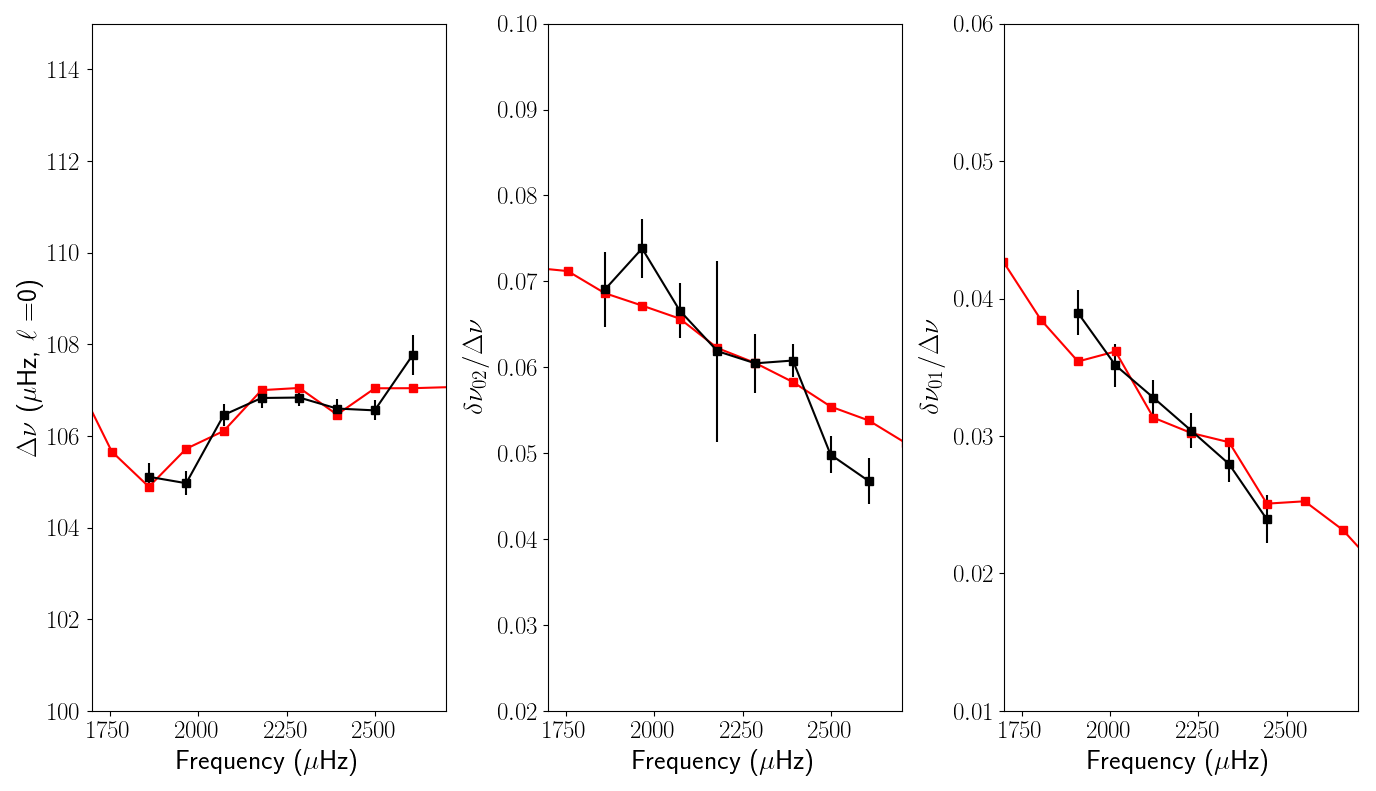}
 \caption{Large separations $\Delta \nu_0$ for $\ell$=0 (left) and small separations $\delta \nu_{02}$ (center) and ratio $\delta_{01}/\Delta\nu_1$ (right) for the star HD43587 modeled with CESTAM. Red lines are for the models and black ones are for the observations, with the associated error bars.}
 \label{fig:separationsHD43587}
\end{figure*}

\begin{table}
\center{
\begin{tabular}{|c|c|c|}
\hline
 & CESTAM & Observed\\
\hline
 Mass ($M_{\odot}$) & $1.04 \pm 0.01$ & - \\
 Radius ($R_{\odot}$) & 1.19 & - \\
 Age (Gyr) & $6.2 \pm 0.1$ & - \\
 $T_{\rm eff}$ (K) & 5966 & $5947\pm17$ \\
 $L$ ($L_{\odot}$) & 1.60 & $1.605\pm0.03$ \\
 $[{\rm Fe/H}]$ & -0.034 & $-0.02\pm0.02$ \\
 $Y_0$ & $0.271\pm0.008$ & - \\
 $(Z/X)_0$ & $0.0261\pm0.0007$ & - \\
 $\alpha$ & $0.692\pm0.005$ & - \\
 $\chi^2_{\mathrm{r}}$ & 3.1 & - \\
\hline
\end{tabular}
\caption{Results from the CESTAM modeling of HD43587.}
\label{tab:res43}
}
\end{table}

The search for the best model was made using different sets of seismic constraints among the ones listed in Sect.\,\ref{sec:Models}, leading to models with optimized parameters that differed by amounts within the error bars. The best (lowest $\chi^2_{\mathrm{r}}$) model matches globally quite well the observed large frequency separation ($\Delta\nu$) and is also in good agreement with the frequency separation ratios $r_{01}$ and $r_{02}$, as well as with observed spectroscopic and photometric values within $1 \sigma$ (see Table \ref{tab:res43}). Fig.\,\ref{fig:HRdiag} presents the HR diagram with the best-fitted model and Fig.\,\ref{fig:separationsHD43587} shows the frequency separations or ratios $\Delta \nu_0$, $r_{02} = \delta \nu_{02}/\Delta\nu$, and $r_{01} = \delta\nu_{01}/\Delta\nu_1$ of this model compared to the observed ones. The differences with observations are quantified as $\chi^2_{\mathrm{r}}=3.1$, which relatively high value can be explained by random differences in seismic differences indicating that the modelling can still be improved. However, the general slopes of the frequency ratios $r_{01}$ and $r_{02}$, which depend on the central hydrogen content, and thus on the stellar age \citep{brandao10,silvaAguirre11}, are correctly reproduced. The present model is also close to the one found by \citet{boumier14} based on a previous version of the stellar evolutionary code CESTAM and a different computation of seismic frequencies. Uncertainties (1$\sigma$ values) are given when the considered parameter is optimized (such as for example the mass and the age). When no uncertainty is given, the parameter considered is not fitted and corresponds to the value obtained in the optimized model. As mentioned in Section\,\ref{sec:Models}, uncertainties are computed using the Hessian matrix of the fitted parameters and do not include other sources of uncertainties. They must be considered as lower bounds. Actual uncertainty interval (accuracy) is larger (roughly estimated to be of at least 0.5\,Gyr for the age), as for example in the case of the age that could be different for modeling using other physical description or other chemical composition \citep[see for instance][]{lebreton&goupil14}.

The value of $\chi^2_{\mathrm{r}}$, larger than unity, indicates a statistically bad agreement. However, as all the inferences of the global parameters ($L$, $T_{\rm eff}$, $[Fe/H]$) are within $1\sigma$ of the observations, these values are mainly due to the difficulty of correctly modeling the internal structure, in particular the core and the base of the convective zone. In particular, the small-scale variations in the ratio $r_{01}$, which are not correctly reproduced, are due to changes in stratification at the base of the outer convective zone \citep{otifloranes05}, and may be better reproduced including convective penetration below the convective envelope \citep{lebreton&goupil12}. 

The analysis of the evolutionary status of HD43587 in the literature is somewhat ambiguous. Because of its large lithium content, it was believed to be younger than the Sun, in contradiction with the flat light curve and the absence of chromospheric activity \citep{baliunas95,schroder12,boumier14}. Our modeling, including seismic constraints, implies that HD43587 is older than the Sun, and we suggest that its large lithium abundance can be due to its slightly larger mass, compared to the Sun, which implies a thinner outer convective zone and thus a shallower mixing underneath, preventing the lithium depletion.\\

\subsection{HD42618}

\begin{figure*}
 \centering
 \includegraphics[width=15cm,height=6cm,angle=0]{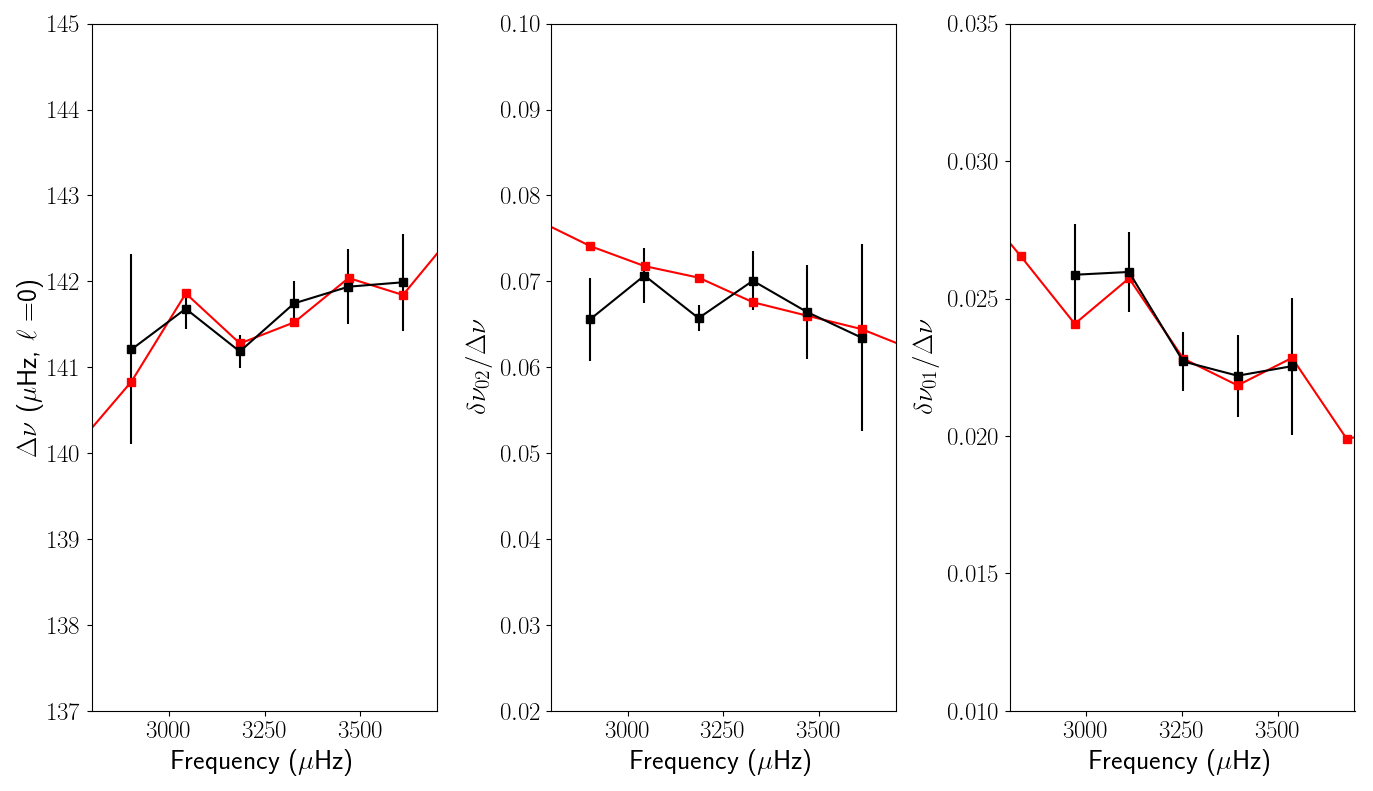}
 \caption{Same as Fig.\,\ref{fig:separationsHD43587}, but for the star HD42618}
 \label{fig:separationsHD42618}
\end{figure*}

As for HD43587, global and seismic constraints listed in Sect.\,\ref{sec:Models} were used for HD42618. Results of the best model are presented in Table\,\ref{tab:res42} and the evolution track is plotted in the HR diagram in Fig.\,\ref{fig:HRdiag} along with the observational point.
The lowest $\chi^2_{\mathrm r}$ model with a value of 0.8 shows a statistically good agreement between observed and modeled seismic large separations and frequency ratios (see Fig.\,\ref{fig:separationsHD42618}) with no systematic differences. Our model reproduces well the shape of the large separations curve, as well as the slope of the frequency ratios, and the small-scale variations of the ratio $r_{01}$. Spectroscopic characteristics are also in very good agreement (less or equal to 2\,$\sigma$ uncertainty). The luminosity of the model is lower, at 3\,$\sigma$ of the observed one. For the metallicity, a difference smaller than 1$\sigma$ is found. 
 
We found a slightly less massive and older star than the Sun. The higher lithium content compared to the solar case can be explained by a significantly lower metallicity, which diminishes the opacity in the outer layers, shallows the depth of the convective zone, and thus the additional mixing beneath it.
 
\begin{table}
\center{
\begin{tabular}{|c|c|c|}
\hline
 & CESTAM & Observed \\
\hline
 Mass ($M_{\odot}$) & $0.92\pm0.02$ & - \\
 Radius ($R_{\odot}$) & 0.94 & - \\
 Age (Gyr) & $5.5\pm0.2$ & - \\
 $T_{\rm eff}$ (K) & $5791$ & $5765\pm17$ \\
 $L$ ($L_{\odot}$) & $0.89$ & $0.918\pm0.012$ \\
 $[{\rm Fe/H}]$ & $-0.116$ & $-0.10\pm0.02$ \\
 $Y_0$ & $0.281\pm0.009$ & - \\
 $(Z/X)_0$ & $0.0206\pm0.0007$ & - \\
 $\alpha$ & $0.686\pm0.011$ & - \\
 $\chi^2_{\mathrm{r}}$ & 0.8 & - \\
\hline
\end{tabular}
\caption{Results from the CESTAM modeling of HD42618.}
\label{tab:res42}
}
\end{table}

\subsection{Chemical clocks}

Recent studies based on high-precision analysis of solar analogues with metallicities near the solar value (-0.15 $\lesssim$ [Fe/H] $\lesssim$ +0.15) have shown remarkably tight and steep correlations between isochrone ages and either [Y/Mg] or [Y/Al]. The age scatter is typically less than 1 Gyr for a given abundance ratio and the relations extend over $\sim$10 Gyr \citep[e.g.,][]{nissen16}. Similar trends are found for stars with asteroseismic ages with uncertainties within 10-20\% ~\citep{nissen17,morel21}. It is believed that the correlations arise from the chemical evolution of the Galaxy \citep[e.g.,][]{spina16}.

We make use of the [Y/Mg] and [Y/Al] abundances of \citet{morel13} and the quadratic age-abundance calibrations of \citet{spina18} to get an independent estimate of the ages of our targets (Table \ref{tab_ages}). We obtain average ages of about 5.4 and 6.3 Gyr for HD42618 and HD43587, respectively. Other calibrations \citep{nissen16,nissen17,spina16,tuccimaia16} lead to younger ages for HD42618, but by less than 0.8 Gyrs. For HD43587, the deviations do not exceed 0.7 Gyr with no evidence of systematic differences. Similar ages are therefore obtained despite the fact that the calibrations rely on different abundance and isochrone datasets.

These results are consistent with our analysis. For both stars, our age estimates lie between the ages inferred from [Y/Mg] and [Y/Al] abundances ratio, with differences around 0.4-0.6 Gyr. In any case, all the ages provided by the chemical clocks are compatible within the error bars.

\begin{table}
\caption{Stellar ages of HD43587 and HD42618 derived from [Y/Mg] and [Y/Al] abundance ratios. The uncertainties in the abundance ratios and calibrations are propagated into the age estimates.}
\label{tab_ages}
\centering
\begin{tabular}{lcc}
\hline
Star    & \multicolumn{2}{c}{Age [Gyr]} \\
        & $[$Y/Mg$]$    & $[$Y/Al$]$    \\
\hline
HD43587 & 6.83$\pm$0.67 & 5.72$\pm$0.66 \\
HD42618 & 5.93$\pm$0.71 & 4.95$\pm$0.74 \\
\hline
\end{tabular}
\end{table}

\section{Conclusions}
\label{sec:conclusions}

In the perspective of the preparation of the PLATO mission \citep{rauer14}, the characterization of solar analogue stars is essential for Earth-like planets hunting. In particular, mass and age are parameters very difficult to estimate, with no direct observation for single field stars. To achieve a better estimate of these parameters, all kind of data are useful, provided they are precise enough. In this work, we used simultaneously spectroscopic data from HARPS and seismic analysis from CoRoT light curves of two solar analogues stars, HD42618 and HD43587. 

A first result concerns HD43587: we found that the star is slightly more massive and older than the Sun. This is in agreement with \citet{boumier14} concerning the mass but we converge to an age larger by about 0.5\,Gyr. This is comparable to actual error bars but could be due to including ratio of frequency separations in the seismic constraints instead of individual frequencies only. However, the relatively high value of the reduced $\chi^2_r$ indicates that our modelling can still be improved.

In the case of HD42618, we converge to an age very different from the estimation of \citet{morel13} based on isochrone fitting (2.17 Gyr but with large error bars of $\pm 1.83$ Gyr), and from \citet{barban13} ($3.84 \pm 0.12$ Gyr) using the Asteroseismic Modeling Portal \citep[AMP,][]{metcalfe09,mathur12}. Our modeling, leading to a low value of the reduced $\chi^2_r$, points clearly to a star slightly more evolved than the Sun but less massive.

We used the [Y/Mg] and [Y/Al] abundances ratio from \citet{morel13}, that show tight correlations with age. These chemical clocks provide age estimates in agreement with our model-inferred ages, improving the reliability of our results.

This work also confirms that to characterize a star (age, mass, radius, etc.), both spectroscopic and seismic measurements must be used, since one or the other of these constraints alone is not enough to guarantee the reliability of the result \citep{piau09,silvaAguirre13,lebreton&goupil14,bazot18}. These stars being finely modeled thanks to seismology, allow a more precise comparison between models and observations. In addition, knowing precisely spectroscopic and astrometric constraints, based on Gaia measurements for example, and other constraints such as rotation or lithium and/or beryllium abundances, in order to assess hypotheses on the efficiency of the internal mixing, are important for future characterization of stars, for example in the frame of the PLATO mission.


\section*{Acknowledgements}
Research activities of the Ge$^3$ stellar team at the Federal University of Rio Grande do Norte are supported by continuous grants from Brazilian scientific promotion agencies. J.D.N. and M.C. acknowledge support from CNPq ({\em Bolsa de Produtividade}). TM acknowledges financial support from Belspo for contract PRODEX PLATO mission development. Funding for the DPAC has been provided by national institutions, in particular the institutions participating in the Gaia Multilateral Agreement. FB warmly thanks Louis Manchon for his help with the handling of CESTAM.


\section*{Data Availability}
This work has made use of data from the European Space Agency (ESA) mission Gaia (\url{https://www.cosmos.esa.int/gaia}), processed by the Gaia Data Processing and Analysis Consortium (DPAC, \url{https://www.cosmos.esa.int/web/gaia/dpac/consortium}), and from the CoRoT public archive (\url{http://idoc-corot.ias.u-psud.fr/}).





\bibliographystyle{mnras}
\bibliography{two_corot_analogues} 


\bsp	
\label{lastpage}
\end{document}